\documentclass[aps,amssymb,floatfix,prd,amsmath,preprintnumbers]{revtex4}
\usepackage{epstopdf}
\usepackage{capt-of}
\usepackage{graphicx}  
\usepackage{dcolumn}   
\usepackage{bm}
\begin{document}
\input epsf.tex
\title{\bf Bulk viscous embedded hybrid dark energy models }

\author{B. Mishra\footnote{Department of Mathematics,Birla Institute of Technology and Science-Pilani, Hyderabad Campus, Hyderabad-500078, India, bivudutta@yahoo.com}, Pratik P Ray\footnote{Department of Mathematics,Birla Institute of Technology and Science-Pilani, Hyderabad Campus,Hyderabad-500078, India, pratik.chika9876@gmail.com}, R. Myrzakulov\footnote{Eurasian International Center for Theoretical Physics and Department of General Theoretical Physics, Eurasian National University, Astana 010008, Kazakhstan, rmyrzakulov@gmail.com
}
}

\affiliation{ }

\begin{abstract}
\begin{center}
\textbf{Abstract}

\end{center}
In this paper, we have constructed the cosmological model of the universe in a two fluids environment with a newly developed mathematical formalism. In order to construct the model Binachi type V (BV) space time is considered with a time varying deceleration parameter. Both the fluids, the viscous fluid and the                    dark energy (DE) fluid have shown their dominance respectively in early time and late time of the cosmic evolution. The scale factor that simulates the cosmic transition based on the value of the bulk viscous coefficient. Within the developed formalism, a general form of the skewness parameters is also obtained as a functional form of the scale factor. The physical parameter of the model such as equation of state (EoS) parameter is also derived and analysed. The state finder diagnostic pair are also obtained to understand the geometrical nature of the model.
\end{abstract}

\maketitle

\section{Introduction}

Several observational studies such as type Ia supernovae (SN Ia)\citep{perl, riess}, baryon acoustic oscillations (BAO)\citep{eisen}, galaxy clustering \citep{seljak}, cosmic microwave background (CMB) \citep{cald1, huang, teg} and weak lensing \citep{dev} confirms the accelerated expansion of the universe. This has posed a challenging theoretical problem to the cosmologists to know the exact reason of the expansion. Therefore, cosmologists studied various energy components of the universe. It has been revealed that the reason behind the accelerated expansion of the universe is due to some form of exotic energy stuff dubbed as DE. From the  recent Planck results, it is observed that DE has occupied almost 70 percent of the total mass energy budget of the universe \citep{ade1, ade2}. It is also observed that DE possesses negative pressure which leads to the increase of rate of expansion of the universe \citep{peeb}. Theoretically, the cosmological constant $(\Lambda)$, once abandoned by Einstein, is put forward as a simplest candidate of DE. However, this cosmological constant is not well defined with respect to the fine-tuning and cosmic coincidence puzzles \citep{cope}. This research triggers cosmologists to study DE models with dynamical DE characterized by an effective EoS parameter (EoS) $\omega_{DE}= \frac{p_{DE}}{\rho_{DE}} \neq -1$ \citep{stein, cald2}.\\

On larger scales, our universe is isotropic and homogeneous. Recently Planck collaboration revealed that this property of isotropic and homogeneity of the universe is well defined  by the $\Lambda$CDM model in the Friedmann-Robertson-Walker (FRW) geometry. However, at low multi-poles the $\Lambda$CDM cosmology shows a poor fit to the CMB temperature power spectrum \citep{ade1, ade2}. This indicates that the isotropy and homogeneity were not the essential features of the early universe. Moreover, the recent Planck data results motivate us to construct and analysed the cosmological models with anisotropic geometry to get a deeper understanding on the the evolution of the universe. In this regard, BV space-time is of fundamental importance since it provides the requisite framework.\\

On the other hand, cosmologists have given a lot of importance to viscous fluid matter which is in contrasts with the traditional approach. In this approach, usually the cosmic fluid remains ideal (non-viscous). Hydrodynamicists suggest that because of the turbulence phenomena, inclusion of viscosity becomes mandatory even in homogeneous space without any limits. There are two viscous coefficients discussed in literature such as the the shear viscosity and bulk viscosity. Shear viscosity is the  dominating one as compared to bulk viscosity \citep{land}. Since a lot of interesting results are available on viscous cosmology on the past universe, the methods from particle physics can also be applied to understand the influence of viscosity on the evolution of universe in homogeneous DE models. Also referring the present observational results for the Hubble parameter and standard Friedman formalism, we may explain the description of the universe back up to the inflationary era, or else we may go to the opposite extreme and analyse the probable ultimate fate of the universe. In early universe, the results of both shear and bulk viscosity were explained by Hogeveen et al. \citep{hoge} using kinetic theory. Also, it is indicated that in early universe the impact of viscosity is very small whereas in future universe the impact is significant. Brevik et al.\citep{brev1, brev2} have investigated viscous cosmology in the early universe for both homogeneous and inhomogeneous EoS and examined the viscosity effects on the various inflationary observables. They have also analysed the viscosity-induced crossing through the quintessence-phantom divide and examined the viscosity-driven cosmological bounces .Since viscosity appears to be an important dissipative phenomena in FRW cosmology, therefore it is  expected that cosmological models embedded with bulk viscosity fluid would produce some remarkable  results in the two fluid situations. Moreover, viscosity embedded cosmological models indicates a substantial contribution of bulk viscosity at the inflationary phase \citep{barr, zimda, pavon}. The bulk viscous driven inflation leads to a negative pressure term, which in process results in repulsive gravity and ultimately became a cause for the rapid expansion of the universe \citep{tripa1,maar, lima}. \\

In mixed fluids environment such as dark fluid matter along with usual ordinary matter (Baryonic matter), a number of literature has motivated the researchers to investigate different models in the back drop of General Relativity with different Bianchi forms. DE models with constant deceleration parameter have been constructed and investigated by Akarsu and Kilinc \citep{akarsu1, akarsu2} for  Bianchi type I and III space time. With a variable Equation of State (EoS) parameter, Yadav et al. \citep{yadav1} constructed BV DE cosmological models where the deceleration parameter was assumed to be constant. Several theoretical two fluids DE models either interacting or non-interacting have been discussed widely in the literature \citep{shey, amir1, amir2, tripa3, suresh}. Mishra et al. \citep{ppr1, ppr2} have constructed DE cosmological models with two non interacting fluid situations such as DE fluid with cosmic string and nambu string. In both the models, they have shown that the models are mostly dominated by Phantom behaviour. In a similar approach of two fluid, DE cosmological models were constructed in different general scale factors \cite{ppr4}. With this motivation, here we have considered the BV space time as
\begin{align} \label{eq1}
ds^{2}= dt^{2}- A(t)^{2}dx^{2}- e^{2 \alpha x} \left[B(t)^{2}dy^{2}+C(t)^{2}dz^{2}\right]
\end{align}
The exponent $\alpha\neq0$ in \eqref{eq1} is an arbitrary constant. The total energy momentum tensor (EMT) in presence of both the viscous and DE fluids can be expressed as,
\begin{equation} \label{eq2}
T_{ij}= T_{ij}^{vis} + T_{ij}^{de},
\end{equation}
where, EMT of barotropic bulk viscous fluid is 
\begin{align} \label{eq3}
 T_{ij}^{vis}=  (\rho + \bar{p}) u_{i}u_{j}- \bar{p} g_{ij},
\end{align}
and EMT of DE fluid is
\begin{align} \label{eq4}
 T_{ij}^{de} \nonumber &= diag[\rho^{de}, -p^{de}_{x}, -p^{de}_{y}, -p^{de}_{z}]\\ \nonumber &= diag[1, -\omega^{de}_{x}, -\omega^{de}_{y}, -\omega^{de}_{z}] \rho^{de}\\  &= diag[1, -(\omega^{de}+ \delta), -(\omega^{de}+ \gamma), -(\omega^{de}+ \eta)]\rho^{de},
\end{align}

Here, $u^{i}$ is the four velocity vector of the fluid in a co-moving coordinate system. $\omega^{de}$ and $\rho^{de}$ are respectively the EoS parameter of the DE fluid and DE density parameter. The skewness  parameters $\delta$ on $x$-axis, $\gamma$ from $y$-axis and $\eta$ on $z$-axis are deviations from the EoS parameter $\omega^{de}$ on these three directions. With these consideration on the parameters, in the subsequent section, we have developed the mathematical formalism of the problem.\\

In section II, the basic equations for  BV space time in presence of viscous fluid and DE fluid are formulated along with the physical and kinematic parameters. The pressure anisotropy is incorporated in three dimensions to obtain the anisotropy in the cosmic fluid. In section III, the scale factor known as hybrid is used to obtain a viable solution. The physical importance of the scale factor also discussed. Moreover. the characteristics of deceleration parameter is presented w.r.t. the hybrid scale factor. The functional form of the skewness parameter and EoS parameter are expressed. Also with the help of skewness parameters, the dynamics of the model are described in section IV. At the end, summaries and results are presented in section V.\\

\section{Mathematical Formalism of the model}
In the two fluid description of EMTs as discussed in the previous section, Einstein's field equations of General Relativity, for the space-time (1) can be calculated as,
 
\begin{align}
& \frac{\ddot{B}}{B}+\frac{\ddot{C}}{C}+\frac{\dot{B}\dot{C}}{BC}-\frac{\alpha^{2}}{A^{2}}=-p+3 \zeta H-(\omega_{DE}+\delta)\rho_{DE} \label{eq5} \\ 
& \frac{\ddot{A}}{A}+\frac{\ddot{C}}{C}+\frac{\dot{A}\dot{C}}{AC}-\frac{\alpha^{2}}{A^{2}}=-p+3 \zeta H-(\omega^{de}+\gamma)\rho_{DE} \label{eq6} \\ 
& \frac{\ddot{A}}{A}+\frac{\ddot{B}}{B}+\frac{\dot{A}\dot{B}}{AB}-\frac{\alpha^{2}}{A^{2}}=-p+3 \zeta H-(\omega^{de}+\eta)\rho_{DE} \label{eq7} \\ 
& \frac{\dot{A}\dot{B}}{AB}+\frac{\dot{B}\dot{C}}{BC}+\frac{\dot{C}\dot{A}}{CA}-\frac{3\alpha^{2}}{A^{2}}=\rho+\rho_{DE} \label{eq8}\\
& 2\dfrac{\dot{A}}{A}-\dfrac{\dot{B}}{B}-\dfrac{\dot{C}}{C} =0 \Rightarrow A^2=k_1BC \label{eq9}
\end{align}
where an over dot represents the derivatives of corresponding field variable with respect to $t$ and in Eqn. \eqref{eq9} $k_1=1$. It can be noted that the product of the field variables $A$, $B$ and $C$ gives the volume scale factor from where the average scale factor can be deduced as $R = V^{\frac{1}{3}}$. If $H_x$, $H_y$ and $H_z$ respectively denotes the Hubble parameter's in the direction of $x$,$y$ and $z$ respectively, then the mean Hubble's parameter $H =\frac{1}{3}\Sigma {H_i}=\frac{\dot{R}}{R}$, where $i=x,y,z$. \\

The proper pressure $p$, in case of barotropic cosmic fluid,  is given as, $p= \xi \rho,$ $(0 \leq \xi \leq 1)$. Moreover, the bulk viscosity related to energy density with the help of Hubble's parameter as $3 \zeta H= \epsilon_{0} \rho$. So, the effective pressure which is the mixture of proper pressure and barotropic bulk viscous pressure can be written as,  $\bar{p} = p-3 \zeta H = (\xi - \epsilon_{0}) \rho= \epsilon \rho$, where, $\epsilon$ can be considered as effective viscous coefficient. Now, replacing pressure terms $(p-3 \zeta H)$ in field equations as $\bar{p}$ and framing field variables in terms of Hubble parameter, we obtain

\begin{align}
& 2 \dot{H}+ 4  \dfrac{(m^{2}+m+1)}{(m+1)^{2}} H^{2}-\frac{\alpha^{2}}{A^{2}} = -\bar{p} - (\omega_{DE}+ \delta)\rho_{DE} \label{eq10} \\
& \left(\dfrac{m+3}{m+1}\right)\dot{H}+ \dfrac{(m^{2}+4m+7)}{(m+1)^{2}} H^{2}-\frac{\alpha^{2}}{A^{2}} = -\bar{p}  - (\omega_{DE}+ \gamma)\rho_{DE} \label{eq11}\\
& \left(\dfrac{3m+1}{m+1}\right)\dot{H}+\dfrac{(8m^{2}+4m+1)}{(m+1)^{2}} H^{2}-\frac{\alpha^{2}}{A^{2}} = -\bar{p} - (\omega_{DE}+ \eta)\rho_{DE} \label{eq12}\\
& \dfrac{(2m^{2}+6m+4)}{(m+1)^{2}} H^{2}-\frac{3 \alpha^{2}}{A^{2}} = \rho + \rho_{DE} \label{eq13}
\end{align}

The energy conservation equation for viscous fluid, $T^{ij(vis)}_{;j}=0$ and DE fluid, $T^{ij(de)}_{;j}=0$ can be obtained respectively as 

\begin{equation} \label{eq14}
\dot{\rho}+3(\bar{p}+\rho)\frac{\dot{R}}{R}=0
\end{equation}

and

\begin{equation} \label{eq15}
\dot{\rho}_{DE}+3 \rho_{DE}(\omega_{DE}+1)\frac{\dot{R}}{R}+ \rho_{DE}(\delta H_x+\gamma H_y+\eta H_z)=0
\end{equation}

From \eqref{eq14}, incorporating the relation between Hubble parameter and scale factor,  we get the energy density for the matter field as,
\begin{align} \label{eq16}
\rho=\frac{\rho_{0}}{\left[ e^{\int{H}.dt}\right]^{3(\epsilon +1)}}
\end{align}
where $\rho_0$ is the integration constant or rest energy density of present time.\\

From the literature, it is evident that bulk viscous fluid has an important role in the study of the recent claim of accelerated expansion of the universe. It is already mentioned that the barotropic bulk viscous pressure includes the contributions both from the usual cosmic fluid and from the coefficient of bulk viscosity $\epsilon$. The contribution from bulk viscosity to cosmic pressure is assumed to be proportional to the rest energy density of universe \citep{tripa1}. From the analysis of $p=\epsilon \rho$, we can note that, in case contribution from bulk viscosity becomes more than the usual perfect fluid pressure then the total effective pressure becomes negative with a negative $\epsilon$. The accelerated expansion in the present epoch is usually attributed to a fluid with negative pressure and hence, it can be thought that the contribution coming from the bulk viscosity is greater than the usual pressure. If the usual pressure from perfect fluid equals to the contribution from cosmic bulk viscosity, then the cosmic fluid in the model behaves like a pressure less dusty universe.  However, the presence of an exotic DE form leads to a negative pressure of the universe which simulates an anti-gravity effect that drives the acceleration. If the time variation of the mean Hubble rate is known, then the rest energy density of the universe can be calculated from \eqref{eq16} for a given value of $\epsilon$. \\

From \eqref{eq13} and \eqref{eq16}, we can retrieve, the DE density as,

\begin{equation} \label{eq17}
\rho_{DE}=\dfrac{2(m^2 +4m +1)}{(m+1)^2} \left( \dfrac{\dot{R}^2}{R}\right)-\dfrac{3 \alpha^{2}}{R^2}- \rho_{0} R^{-3(\epsilon + 1)}
\end{equation}

With the help of the second part of  the conservation equation \eqref{eq15},which corresponds to the deviation of equation of the state parameters and other is deviation free part and incorporated value of $\eta$ from \eqref{eq12}, we  formalize the EoS parameter of DE as,

\begin{align}\label{eq18}
\omega_{DE}= -\dfrac{1}{\rho_{DE}} \left[ \dfrac{2(m^{2}+4m+1)}{3(m+1)^{2}} \left( F(R)- 3 \dfrac{\dot{R}^{2}}{R^{2}} \right) - \dfrac{\alpha ^{2}}{R^{2}} + \epsilon \rho_{0} R^{-3(\epsilon +1)} \right]
\end{align}

where, $F(R)= \frac{\ddot{R}}{R}+ 4 \frac{\dot{R}^{2}}{R^{2}}$. With the help of eqns. \eqref{eq17}-\eqref{eq18}, eqns.  \eqref{eq10}-\eqref{eq12} can be formulated as:

\begin{align}
\delta = & - \left( \frac{m-1}{3 \rho^{de}} \right) \chi(m)  F(R) \label{eq19}\\
\gamma = & \left(\frac{5+m}{6 \rho^{de}} \right) \chi(m) F(R) \label{eq20}\\
\eta =& - \left( \frac{5m+1}{6 \rho^{de}} \right) \chi(m) F(R) \label{eq21}
\end{align}

Where, $\chi(m)= \dfrac{m-1}{(m+1)^2}$.

\section{Solution of the model using Hybrid scale factor}
Based on the recent outcomes on the present universe, mostly the scale factors are chosen to be either the exponential law expansion universe or power law expansion, whose deceleration parameter turns out to be constant. However, the time dependence of the directional scale factor would be decided by specific choices of scale factors. In the present work, we have considered the specific scale factor, the hybrid scale factor which at late time results into a constant deceleration parameter. The hybrid scale factor has two factors in the form, $R=e^{at}t^{b}$, where $a= \left( \frac{m+1}{2} \right) \xi$ and $b= \left( \frac{m+1}{2} \right)n$ are positive constants. The cosmic dynamics is dominated by the power law $(t^{b})$in the early phase, whereas it is dominated by the exponential factor $(e^{at})$ at late phase. However, eventually, the hybrid scale factor is found to be more dominant in the late phase of the evolution. It can be noted that in the hybrid scale factor, when the exponent $a=0$ and $b=0$, it recovers power law and exponential law respectively. For this model, the Hubble parameter and the directional Hubble parameter can be obtained respectively as $H = H_{x} = a +\frac{b}{t}$, $H_{y}=\frac{2m}{m+1} (a +\frac{b}{t})$ and $H_{z}=\frac{2}{m+1} (a +\frac{b}{t})$. Hence, with the hybrid scale factor the energy density of the matter from \eqref{eq16} would be $\rho=\frac{\rho_{0}}{\left[ e^{\int{H}.dt}\right]^{3(\epsilon +1)}}= \rho_{0} (e^{\xi t}t^{n})^{-\frac{3}{2}(m+1)(\epsilon +1)}$. Subsequently, DE density and the effective EoS parameter can be written as,

\begin{align} \label{eq22}
\rho_{DE}= \dfrac{(m^{2}+4m+1)(n+ \xi t)^{2}}{2t^{2}}-\dfrac{3 \alpha^{2}}{(e^{\xi t}t^{n})^{(m+1)}}-\dfrac{\rho_{0}}{(e^{\xi t}t^{n})^{\frac{3}{2}(m+1)(\epsilon+1)}}
\end{align}

and  
\begin{align} \label{eq23}
-\omega_{DE} \rho_{DE}= -2(m^{2}+4m+1) \Phi (t)+ \alpha^{2} (e^{\xi t}t^{n})^{-(m+1)} - \epsilon \rho.
\end{align}

where $\Phi(t)= \left[\frac{\xi^2t^2+n^2-2\xi nt}{4t^2}-\frac{n}{3(m+1)t^2}\right]$. Similarly, the skewness parameters can be expressed as,

\begin{align}
\delta = & - \left(\frac{m-1}{\rho^{de}}\right)(m-1)f(t)
 \label{eq24}\\ 
\gamma = & \left(\frac{5+m}{2\rho^{de}} \right)(m-1)f(t) \label{eq25}\\ 
\eta =& - \left( \frac{5m+1}{2 \rho^{de}} \right)(m-1)f(t) \label{eq26}
\end{align}
Here, $f(t)=\left[\Phi(t)+\left(\frac{\xi t+n}{2t}\right)^2\right]$.

It is observed that both $\rho$ and $\rho_{DE}$ are decreasing when time period is gradually increasing. The decrease in $\rho_{DE}$ is decided by three different factors i.e. $\frac{1}{t^{2}}$ , $\frac{1}{t^{n(m+1)}}$ and $\frac{1}{t^{\frac{3n}{2}(m+1)(\epsilon + 1)}}$. However, only the third term contains the bulk viscous coefficient for the DE density. When, $\epsilon=-1$,  the contribution from cosmic fluid for the DE density becomes time invariant. Similarly whenever  $\epsilon=-\frac{1}{3}$, the denominator of second and third term of \eqref{eq22} becomes same and the rest energy density along with the constant $\alpha$ decides the behaviour of DE density. Since bulk viscous coefficient $(\epsilon)$ acts as EoS parameter $(p=\epsilon \rho)$ for viscous source of matter, the above two values of $\epsilon$ reduces the model to vacuum and radiation dominated respectively in absence of DE. So, in order to understand the behaviour of the matter throughout, we have chosen a value $\epsilon=-\frac{2}{3}$, which is in the range $[-1,-\frac{1}{3}]$. Consequently, for this choice, in spite of the presence of the time factor $f(t)$,the skewness parameters become constant and independent of time, so the change in the isotropic pressure in all directions remain constant. To investigate some more details on the DE model, we have derived some physical parameters of the model. The scalar expansion and shear scalar are respectively given by,
\begin{align}
\theta=3H=3\left(a+\frac{b}{t}\right)
\end{align}
\begin{align}
\sigma^2=\frac{1}{2}(H_x^2+H_y^2+H_z^2-\frac{\theta^2}{3})=\biggl(\frac{m-1}{m+1}\biggr)^2\biggl(a+\frac{b}{t}\biggr)^2       
\end{align}
The anisotrpic parameter can be expressed as,
\begin{equation}
\mathcal{A}_m=\frac{1}{3}\sum \biggl(\frac{\vartriangle H_i}{H}\biggr)^2=\frac{2}{3}\biggl(\frac{m-1}{m+1}\biggr)^2
\end{equation}
Here, the parameter $m$ dealt with the anisotropic behaviour of the model and became isotropic for $m=1$. The parameters $H, \theta$ and $\sigma ^{2}$ start with with an extremely large values and continue to decrease with expansion of universe; whereas the spatial volume grows with cosmic time which mimic the present scenario of the universe. 

\section{Dynamical Behaviour of the model}

The deceleration parameter $q=- \frac{R \dot{R}}{\dot{R}^2}$  describes the cosmic dynamics of universe. Positive value of it indicates decelerating universe where as negative values confirms the accelerated expansion of the universe. In view of the observations of high red shift supernova, Type Ia supernova observations combined with BAO and CMB, models transiting from early decelerating universe to late time accelerating universe gained much importance in recent times. According to recent observational data at present time, the most favourable value for $q$ to be $ -0.81 \pm 0.14$. The deceleration parameter for the hybrid scale factor, turns to be $q= -1 + \dfrac{2n}{(m+1)(\xi t+ n)^{2}}=-1 + \dfrac{b}{(at+b)^{2}}$. So, at an early phase of cosmic evolution, i.e. whenever $t\rightarrow 0$, $q\rightarrow -1+\dfrac{1}{b}$ whereas  at late phase i.e. whenever $t\rightarrow \infty$, $q \simeq -1$. The parameter $b$ is constraint here to be in range $0<b<\frac{1}{3}$ to get a transient universe \cite{mishra2}. In Fig-1, we have represented the deceleration parameter with cosmic time. At early phase the deceleration parameter is positive decreases rapidly and at late phase it appears to be negative. At present time, the deceleration parameter value found to be $(\simeq- 0.9)$, which is in alignment with the observational data. \\

\begin{figure}[h!]
\minipage{0.40\textwidth}
\centering
\includegraphics[width=65mm]{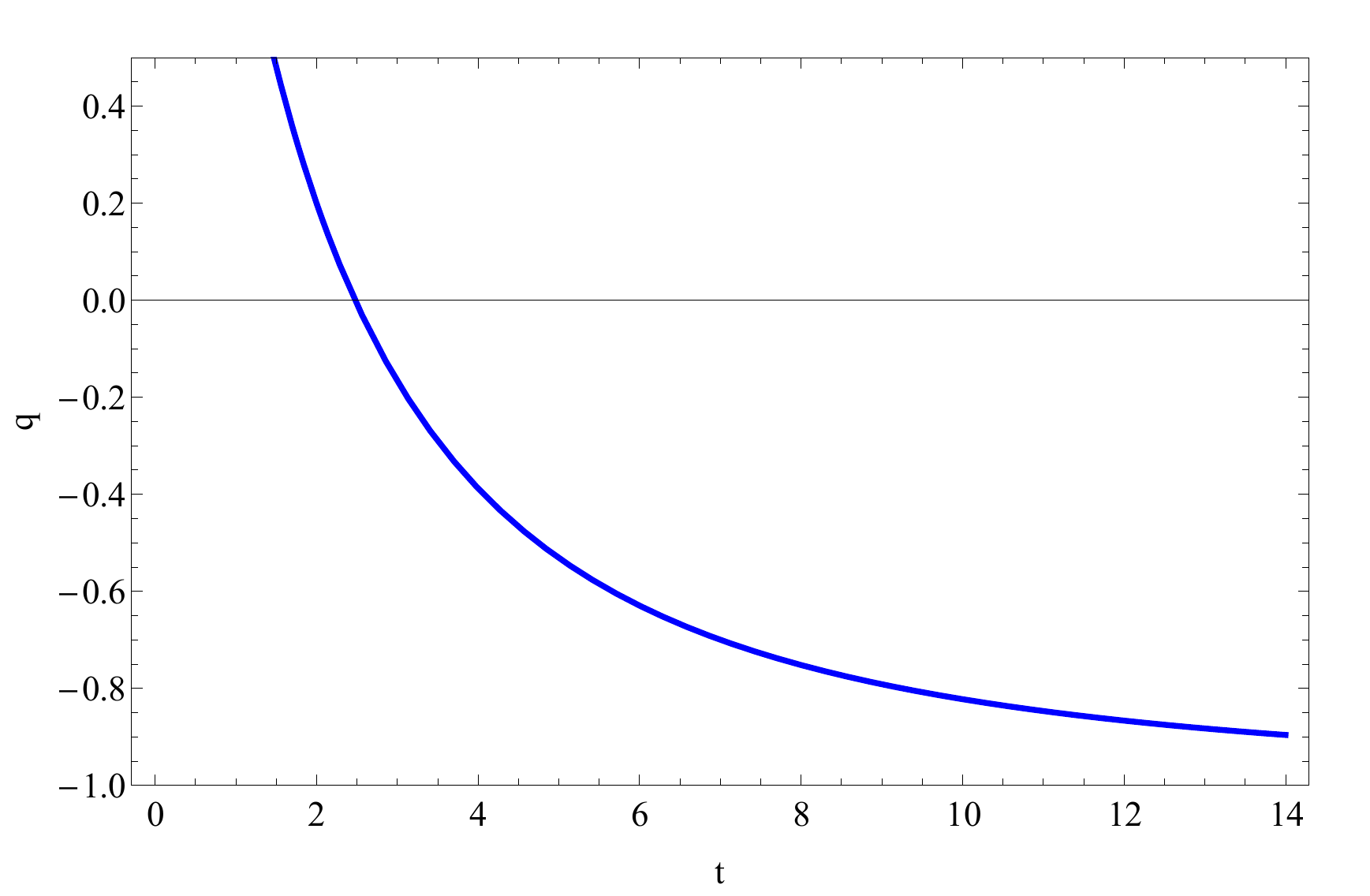}
\caption{DE deceleration parameters  vs. time }
\endminipage\hfill
\minipage{0.40\textwidth}
\includegraphics[width=65mm]{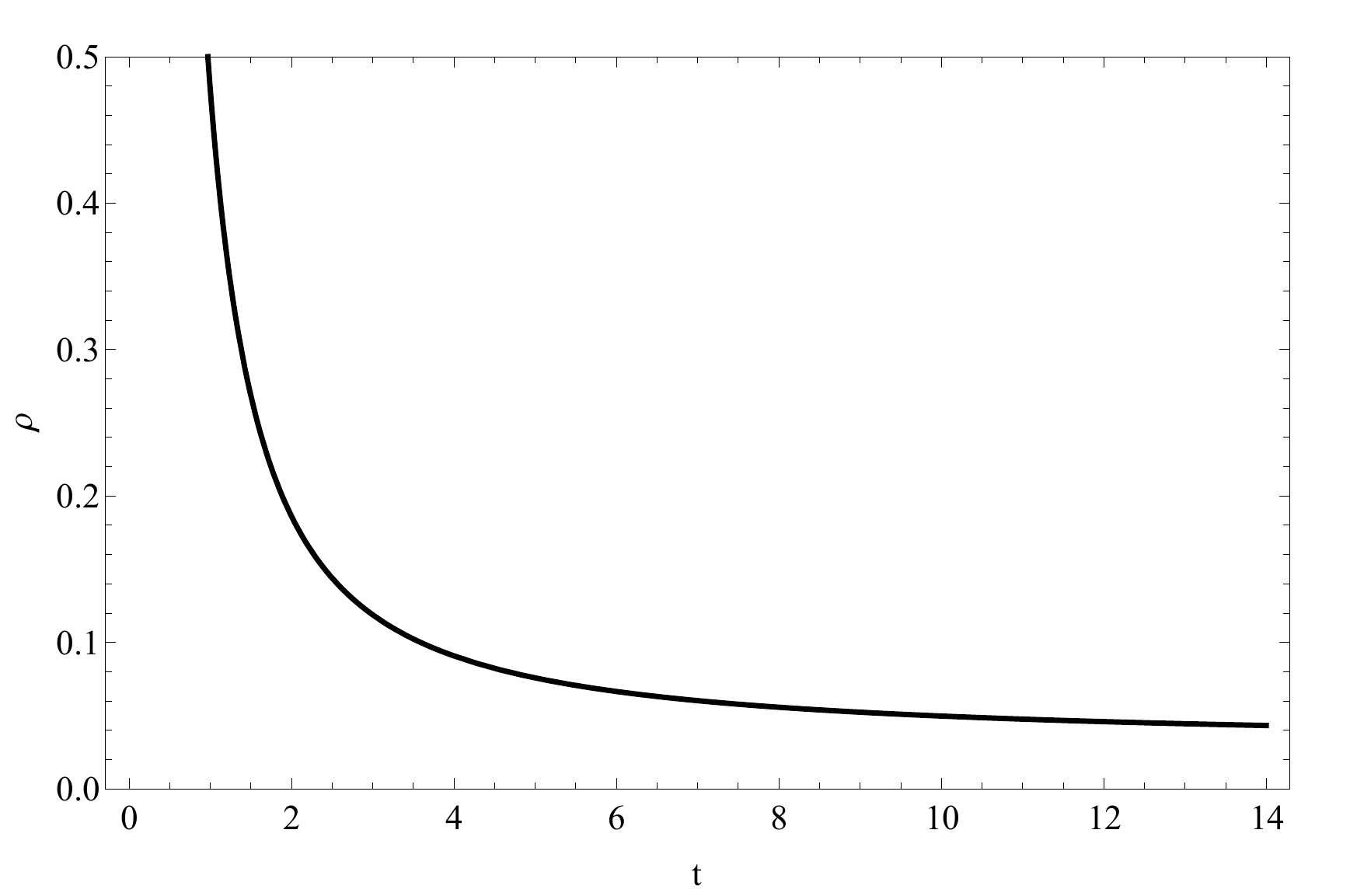}
\caption{DE density parameter  vs. time  }
\endminipage 
\end{figure}

In Fig. 2, we observe, the DE density $(\rho_{DE})$ remains positive till late phase of evolution, satisfying weak energy condition (WEC) and  null energy condition (NEC) for the present model. It is worth to mention that, behaviour of DE density does not depend on change in value of viscous coefficient$(\epsilon)$ and remains alike. $\rho_{DE}$ decreases with increase in cosmic time and reaches to a positive value at present epoch instead of coming closer to zero as in case of the hybrid model embedded in string fluid \cite{ppr1}. This indicates that viscous fluid has smaller effect on $\rho_{DE}$ than string fluid. However, this small effect due to mix fluid matter cannot be ruled out.\\

\begin{figure}[h!]
\minipage{0.40\textwidth}
\centering
\includegraphics[width=65mm]{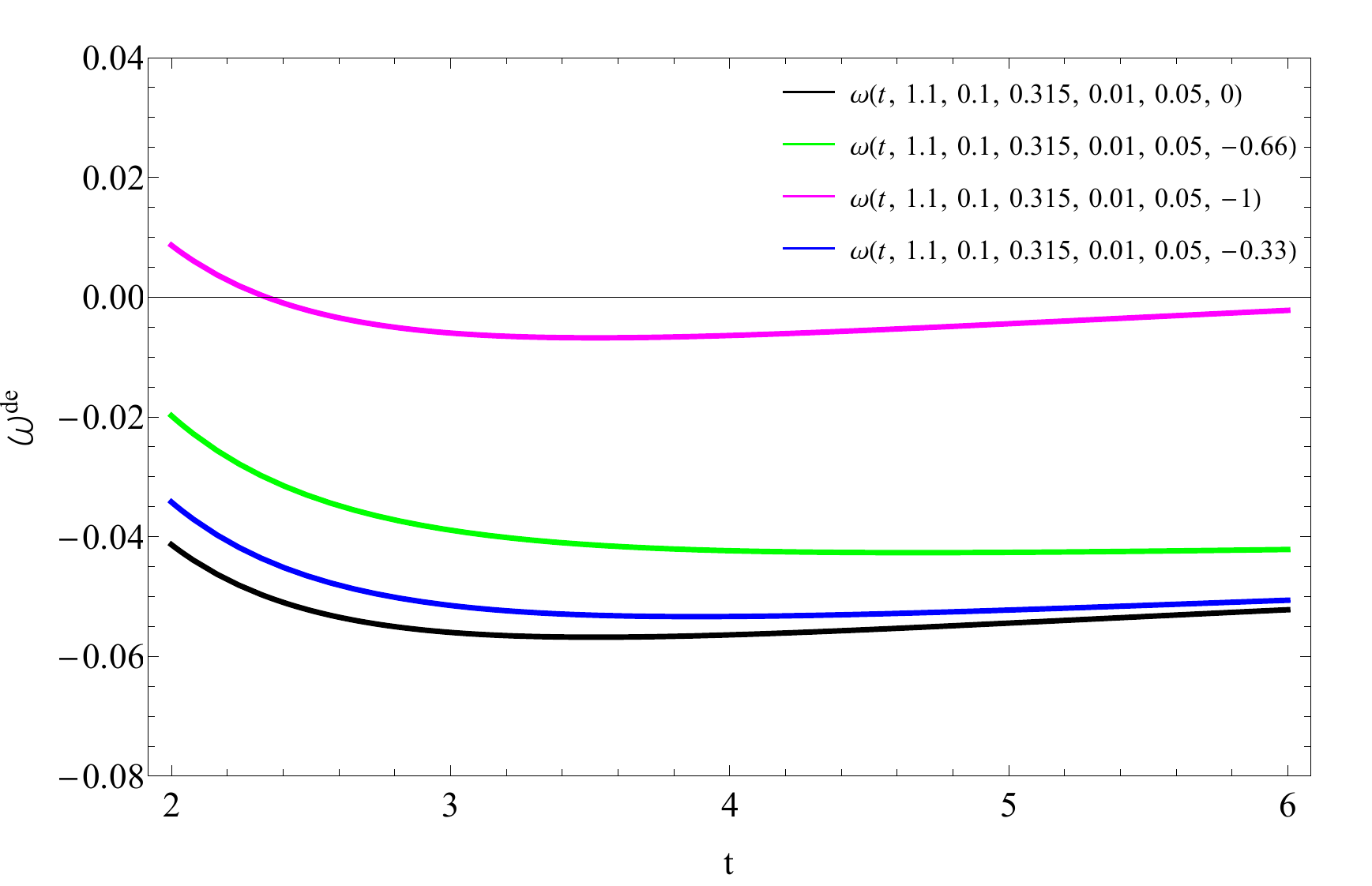}
\caption{DE EoS parameter  vs. time for different viscous coefficients }
\endminipage\hfill
\minipage{0.40\textwidth}
\includegraphics[width=65mm]{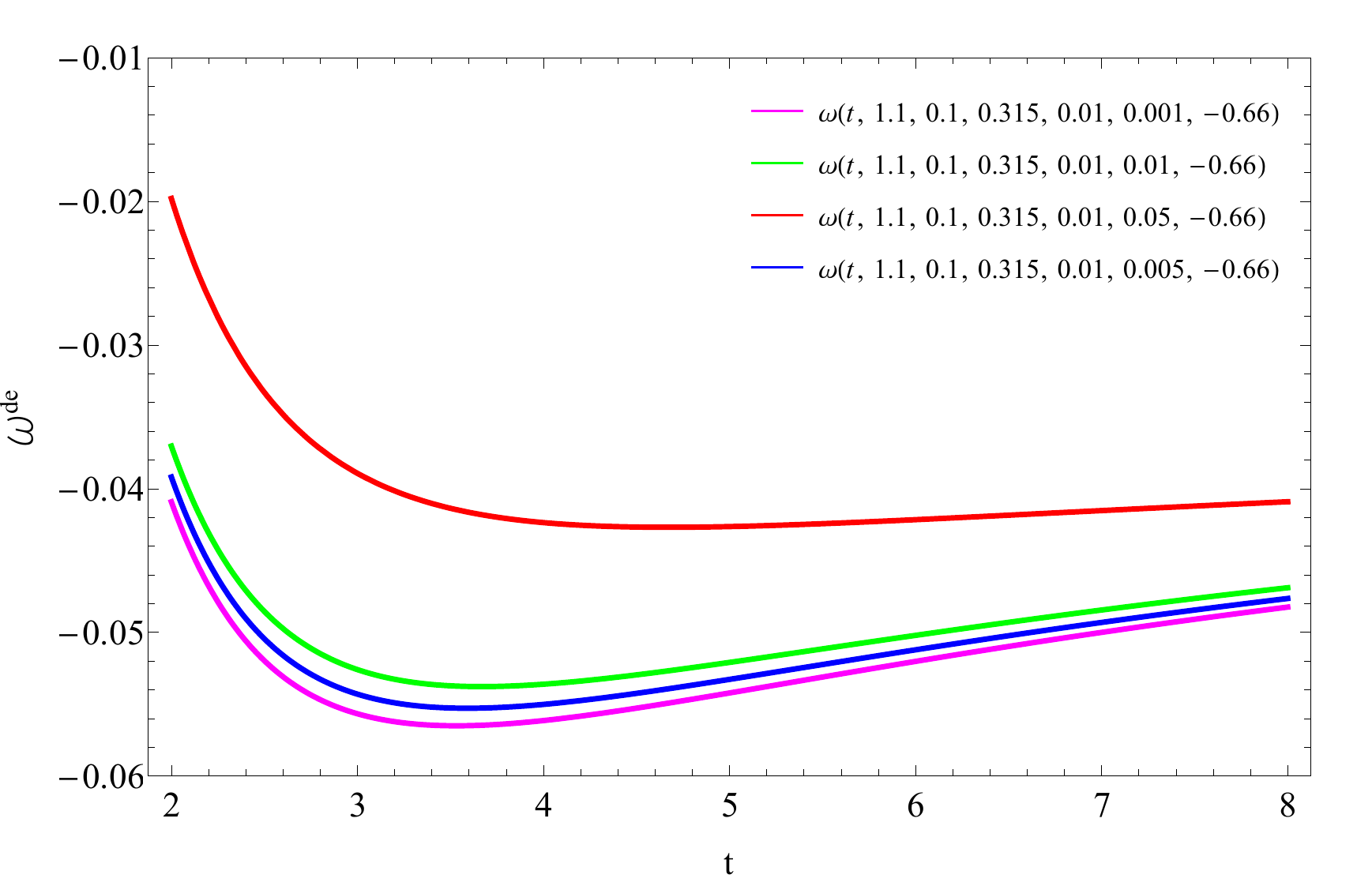}
\caption{DE EoS parameter  vs. time for different $\alpha$ with $\epsilon=-0.66$  }
\endminipage 
\end{figure}

\begin{figure}[h!]
\minipage{0.40\textwidth}
\centering
\includegraphics[width=65mm]{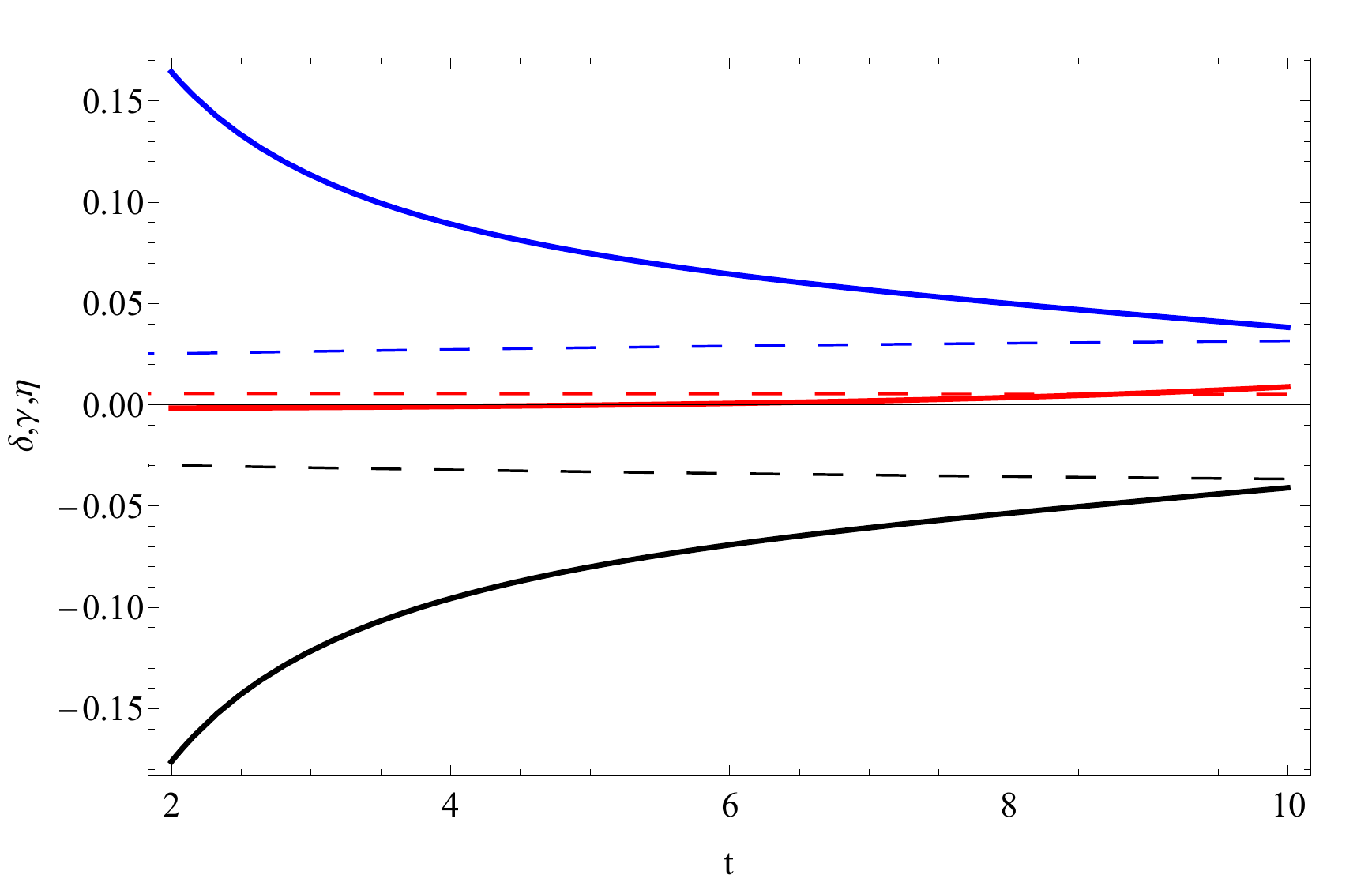}
\caption{DE skewness parameters  vs. time }
\endminipage\hfill

\end{figure}

Fig.3 represents variation of EoS parameter $(\omega_{DE})$ w.r.t appropriate choices of bulk viscous coefficients $(\epsilon= 0, -\frac{1}{3}, -\frac{2}{3}, -1)$. When compared to the de Sitter model and power law model \cite{ppr3}, we found nature of EoS parameters are directly proportional to the increasing value of viscous coefficients $(\epsilon)$ till late phase. But the behaviours of the parameters greatly affect the dynamics at early phase of evolution where as the late phase is smooth and mildly affected by the viscous coefficient values, falling in the preferred range, determined by observational data. The reason of this drastically affected early phase is due to the substantial contribution of bulk viscous fluid. The pink line $(\epsilon=-1)$ confirms that the less is the viscous coefficient value, lesser is the possibility of occurring $\omega_{DE}$ within the observed length (Quintessence region). However, for much lower value of $(\epsilon)$ (beyond $\epsilon=-1$), the decrement of EoS parameter is more rapid. Also, $\omega_{DE}$ decreases with increase in cosmic time in these cases. With increase in viscous coefficient value, this model gathers some energy in early phase and behave differently. The black line $(\epsilon=0)$ represents the cosmic fluid without usual matter but only of dark fluid. This line is most closer to $\Lambda$CDM line than other EoS parameter lines which are embedded with viscous fluid. It also indicates that without any viscous fluid, the model behaves like quintessence field with cosmic growth. The green line $(\epsilon=-\frac{2}{3})$ and blue line $(\epsilon=-\frac{1}{3})$ emerge at early phase, showing little deflection due to presence of viscous fluid and decrease smoothly in the quintessence region with evolution. At late phase of cosmic evolution, $\omega_{DE},$ for all the cases considered here except $(\epsilon=-1),$ decrease to achieve larger negative value. In spite of the presence of bulk viscous fluid, the DE seems to be dominant the universe. Hence, it indicates that there is a very little impact of bulk viscous on the dynamics of EoS, that too in the early phase of evolution.\\

DE EoS parameters can also be useful in testing the model w.r.t certain other parameters. In fact, one has the liberty to test the impact of EoS parameter for different choices of parameter $\alpha$, shown in Fig. 4. $\omega_{DE}$ remains in the quintessence region for all considered values of $\alpha$ but the behaviour changes over time and inclined towards $\Lambda$CDM line. Behaviour of $\omega_{DE}$ is same as compared to different viscous coefficient values. But when the value of $\alpha$ increases, $\omega_{DE}$ increases most rapidly. Also, the value of $\omega_{DE}$ for $\alpha=0.05$(red line) is more for higher values of $\alpha$ and lies very similar to $\omega_{DE}$ for more suitable viscous coefficient value $(\epsilon= -\frac{2}{3}).$\\

In absence of cosmic bulk viscous fluid (dotted lines), the anisotropic parameters execute almost non-evolving behaviour in most of the early phase evolve a little towards late phase \cite{ppr3}. The DE pressure along x- direction,  remains unaffected mostly during evolution. But the anisotropy in DE pressure along y-direction, $\gamma$(blue dotted line) increase a little, where as, along z-direction, DE pressure $\eta$(black dotted line) decrease a little at late times. One can also conclude that, at initial phase of phase of evolution, the skewness parameter may merge into one line indicating the isotropic universe but at late phase is anisotropic w.r.t small scale contribution. In presence of cosmic bulk viscous fluid, the early phase is dominated and affected mostly by bulk viscous but remain unaltered as in case of no viscous fluid, towards$\delta$(red dotted line) late phase. $\gamma$(blue solid line) and $\eta$(black solid line) show similar increasing and decreasing behaviour, being affected by viscous fluid. In fact, they both evolve as mirror image to each other in both presence and absence of cosmic viscous fluid. However, the viscous fluid affects the skewness parameter $\delta$(red solid line) mostly as compared to the other two skewness parameters. $\delta$ increases as it moves towards late phase and changes sign at some cosmic time. This impact on $\delta$ may be due to the fact that, we have considered the mean Hubble parameter is same as directional Hubble parameter along x-axis $(H=H_{x})$. Also, the effects of anisotropic parameter ($m$), parameters $\alpha$ and $n$ are investigated w.r.t pressure anisotropy. But they do not find to have any significant impact on the overall behaviour of skewness parameters.\\

The geometrical behaviour of the DE model can be assessed through the state finder diagnostic  pair $(r,s)$. The acceptability of corresponding DE Hybrid (DEH)  model can be decided through the $(r, s)$ diagnosis comparing with the standard $\Lambda CDM$ model. Hence, we have analysed the evolutionary behaviour of both the parameters $r$ and $s$ for the DE universe along with $\Lambda CDM$ universe. Both parameters evolve continuously with time from big bang time $(t\rightarrow 0)$ to large value at late time $(t\rightarrow \infty)$. The pair can be obtained as,

$$r= 1-\frac{6n}{(m+1)(\xi t+n)^2}+\frac{8n}{(m+1)^2(\xi t+n)^3}$$

$$s=-\frac{12(m+1)(\xi t+n)n+16(m+1)n}{12(m+1)(\xi t+n)n-(m+1)^2(\xi t+n)^3} $$

Here, $q$ and $r$ are respectively be the deceleration parameter and jerk parameter. Parameter $s$ is introduced to characterize the property of DE. The values of $(r,s)$ depend on the anisotropic parameter $(m)$ and constant $(n)$ of the hybrid scale factor chosen. At early cosmic time (big bang time), the state finder pair can be calculated as $\left(1-\frac{6n(m+1)+8}{(m+1)^2n^2},-\frac{12n+16}{12n+(m+1)n^2}\right)$. At the later phase cosmic evolution, the model obtained here behaves as $\Lambda CDM$ model, with $(r,s)$ values to be $(0,1)$. Fig. 6 describes the state finder diagnosis on $rs$- plane. Horizontal and vertical lines intersect at $\Lambda CDM$ point through which the curve passes along with cosmic evolution. It ensures that our considered DEH model is well acceptable. Also, the present value of $(r,s)$, indicated by a black dot in the figure, is in nice agreement with recent observational data \cite{ade2}. \\

In Fig. 7, the evolutionary behaviour of the present DEH model and $\Lambda CDM$ model are plotted in the $rq$- plane. For similar kinematics, this is an effective way to compare and differentiate different cosmological models. Vertical lines stand for different eras of cosmic evolution, starting from BBN( big bang nucleosynthesis) to de-Sitter state $(q \thicksim -1)$. It is observed that the present DEH model evolves from a radiation dominated era to the de Sitter phase. Several different models have different evolution trajectories whereas evolutionary behaviours in those models remain in the range $q \lesssim 0.5$, approaching to same future (de Sitter universe). But in the DEH model developed here, the universe may be described from the primordial nucleosynthesis time of universe \cite{akarsu3}. According to general relativity the evolution of the universe start from dust dominated era ($q = 0.5$ and $ r = 1$) whereas in our presented DEH model the universe starts evolving from the BBN times.

\begin{figure}[h!]
\minipage{0.40\textwidth}
\centering
\includegraphics[width=30mm]{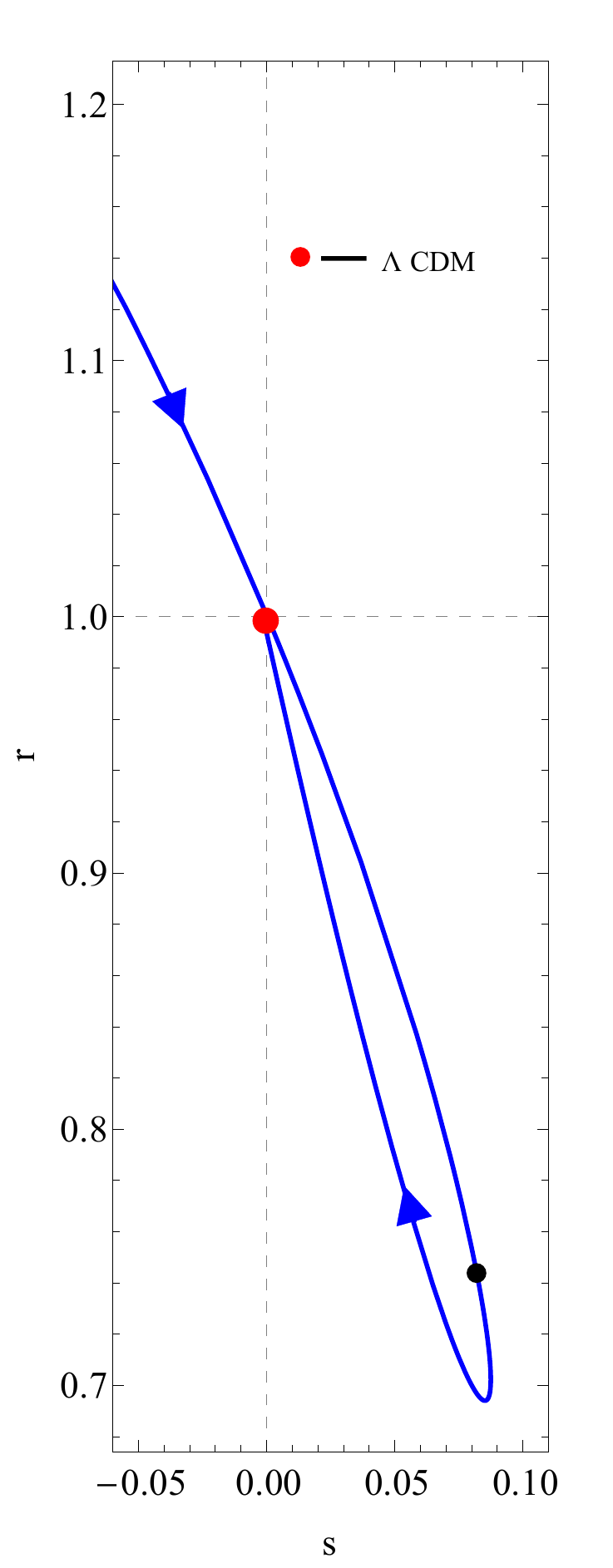}
\caption{r vs s}
\endminipage\hfill
\minipage{0.40\textwidth}
\includegraphics[width=75mm]{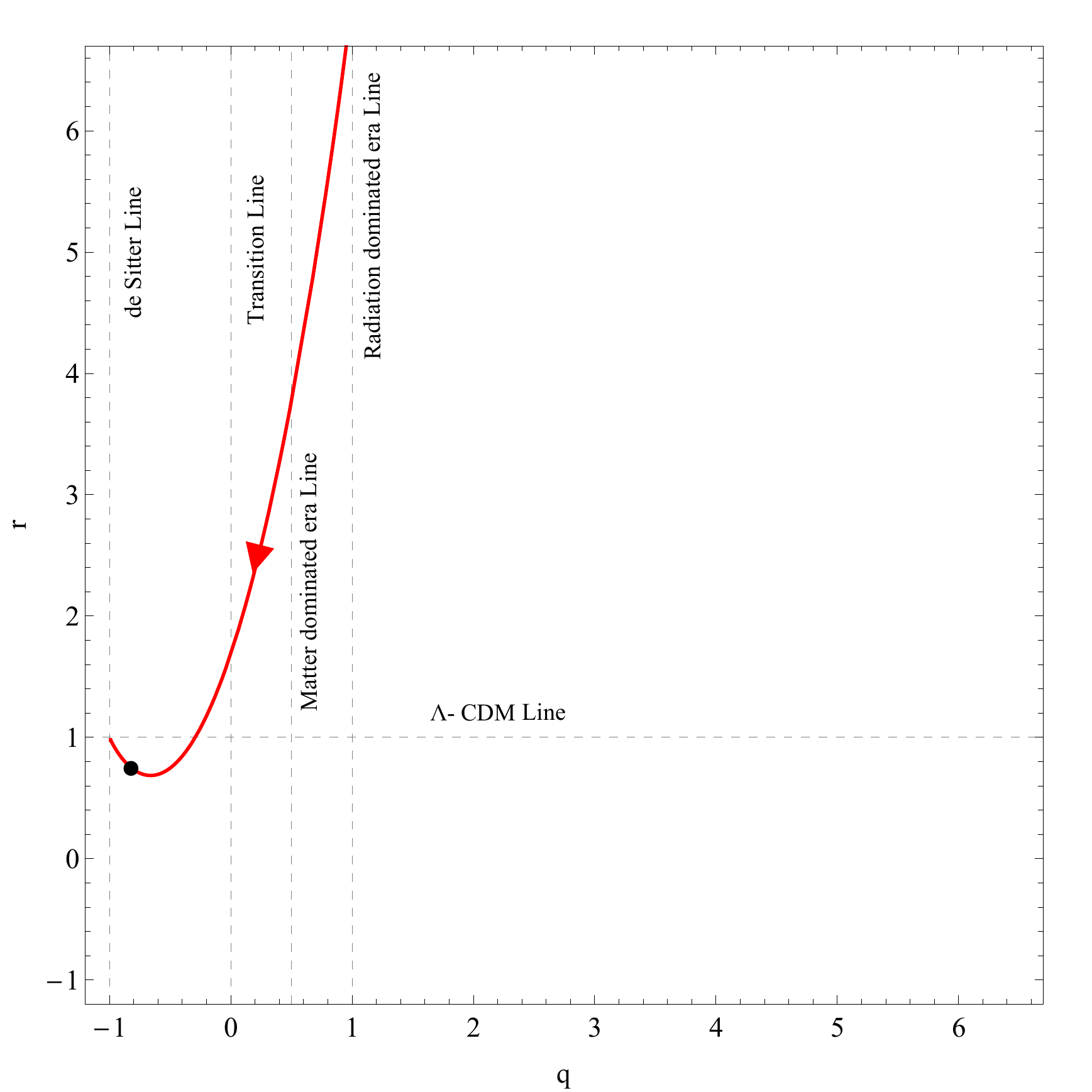}
\caption{r vs q}
\endminipage 
\end{figure}

\section{Conclusion}

In this work, we have investigated the anisotropic behaviour of the cosmological model constructed in a two fluid situations: the usual bulk viscous fluid and DE fluid. The scale factor considered here is the hybrid scale factor which can be attributed to power law cosmology and de Sitter universe for appropriate value of the constant. The parameters of the scale factor has been chosen appropriately from some physical background. Along, $x$-direction, the anisotropy in DE pressure has a very little effect on the pressure anisotropy  whereas along y-direction, it increases and along z-direction decreases at late times. Presence of viscous fluid substantially affects the DE density at early phase of cosmic evolution; however at late phase late phase DE density dominates over viscous fluid. One more observation is that, the DE EoS parameter remains unaffected for different viscous coefficients however it for different value of the constant $\alpha$, it remains in the quintessence region. The anisotropic parameters remain almost non-evolving in most of the early phase in the absence of viscous fluid however in the presence of viscous fluid it has impact. In the late phase the DE fluid has the dominance over bulk viscous fluid. From the evolutionary behaviour of the state finder pairs, we can also conclude that the model behaves like $\Lambda CDM$ universe.

\end{document}